\documentclass[%
 ,secnumarabic%
,amssymb, amsmath,nobibnotes, aps, prl,
showpacs,showkeys]{revtex4}
\usepackage{docs}%
\usepackage{bm}%
\usepackage{graphicx}
\expandafter\ifx\csname package@font\endcsname\relax\else
 \expandafter\expandafter
 \expandafter\usepackage
 \expandafter\expandafter
 \expandafter{\csname package@font\endcsname}%
\fi


\def\GeV{{\,\mathrm{GeV}}}
\begin{document}

\title{Quark-hadron duality in the Rein-Sehgal model}%

\author{Krzysztof M. Graczyk}
\email{kgraczyk@ift.uni.wroc.pl}
\author{Cezary Juszczak}
\author{Jan T. Sobczyk}
\affiliation{ Institute of Theoretical Physics, University of Wroc\l
aw, pl. M. Borna 9, 50-204, Wroc\l aw, Poland}
\date{\today}%

\begin{abstract} The quark-hadron duality in CC and NC neutrino interactions
    is discussed under assumptions that single pion production is described
    accurately by the Rein-Sehgal model and that it allows reconstruction of
    the inclusive cross section in the resonance region. The duality is measured by
    means of integrals of structure functions in the Nachtmann
    variable for $Q^2<3$ GeV$^2$. The results depend on the precision with which
    contributions
    from single pion production channels in the overall cross sections are
    known. Several approaches to evaluate them are compared.
The duality is predicted to be seen for proton
    target reactions and to be absent for neutron and isoscalar
    targets. Two-component duality between resonant and valence quark contributions to
    structure functions is also investigated.
\end{abstract}

\maketitle
%\tableofcontents

\section{Introduction}
In recent years the quark-hadron (QH) duality has been a subject
of many experimental and theoretical studies. The idea of duality
comes from Bloom and Gilman \cite{BG}. It was based on the
comparison of $eN$ $F_2$ resonance structure functions at small
values of $Q^2$ with their DIS counterparts at large $Q^2$. Bloom
and Gilman observed that in plots in the variable
$(2M\nu+M^2)/Q^2$ resonance peaks are approximately averaged by
the DIS structure function. More recent experimental data on
duality were reported in \cite{Nic} where several plots of $F_2$
for $Q^2$ in the range from 0.3 to $3.3~\mathrm{GeV}^2$ as
functions of the Nachtmann variable $\xi$ were presented on the
same figure. The resonance peaks for $\xi \geq 0.2$ are seen to be
averaged by the DIS structure function calculated at
$Q^2=5~\mathrm{GeV}^2$. The QH duality can be analyzed in
quantitative way by calculating the ratios of integrated strengths
over a range in $\xi$ covering a chosen set of resonances and the
DIS structure function. The agreement is on the level of 10\%.

In the theoretical analysis of the duality \cite{MEK} one
introduces a language of twist expansion of moments of structure
functions in powers of $1/Q^2$ . The duality then means the
suppression of higher twists \cite{Ru}. A possible explanation how
the coherent amplitude in the resonance region can be equal to
incoherent sum of amplitudes from the quark constituents of
nucleon is proposed in \cite{CI} in terms of SU(6) symmetry by
means of cancellation of contributions of positive and negative
parity. The knowledge of relative strengths of electromagnetic and
neutrino induced $N\rightarrow N^*$ transitions leads to
theoretical predictions concerning the domain in $W$ in which the
duality should hold. The consequences of SU(6) breakdown for
ratios of unpolarized and polarized structure functions are also
investigated in \cite{CM}.

The QH duality is usually discussed theoretically and analyzed
experimentally in the context of eN interactions. But the subject is
relevant also for neutrino physics. Here the exact data is missing
and arguments based on duality can be used to provide better
estimates of $\nu$N structure functions and cross sections in the
few GeV neutrino energy region where they are known with
insufficient accuracy \cite{NuInt}. For these energies it is
necessary to consider both quasi-elastic and inelastic channels.
Single pion production (SPP) channels are typically treated
separately from more inelastic ones which are accounted for by
extrapolating DIS formalism as much as necessary. This approach
carries a lot of uncertainty and requires a better theoretical
understanding in the whole kinematical region. Such understanding
and control over the numerical procedures can follow from the QH
duality analysis.

When trying to discuss the QH duality in the neutrino interactions
the main obstacle is a lack of precise experimental results. The
existing SPP data is poor in precision and statistics \cite{Zel}.
In the future the data will hopefully become precise enough to
impose more rigid constraints on theoretical models \cite{MIN} but
for a moment the natural strategy is to analyze in detail a
generally accepted model. From the point of view of Monte Carlo
application this is the Rein-Sehgal (RS) \cite{RS} model.

The RS model was developed to describe SPP induced by lepton-nucleon
NC and CC interactions. It is currently used in almost all neutrino
Monte Carlo generators of events \cite{NuInt}. It is based on the
quark model computations of the hadronic current as proposed and
developed by Feynman, Kislinger and Ravndal \cite{FKR}. In the
original model the contributions from 18 resonances with masses
smaller than 2~GeV are added in the coherent way. The scattering
amplitudes contain several form factors. The vector form factors are
obtained from the electro-production data by means of standard CVC
argument. Axial form factors are less constrained by theoretical
arguments (PCAC) and the available data. The RS model contains also
an {\it ad hoc} prescription for computing the non-resonant
background. The extra amplitude is introduced with the quantum
numbers of the $P_{11}$ resonance but without the Breit-Wigner term
and then added in the incoherent way. Its strength is fine tuned in
order to obtain an agreement with the experimental SPP data. The RS
model can be supplemented by $m^2$ containing terms \cite{Nau} ($m$
is the charged lepton mass) absent in the original paper.

The discussion of the QH duality for neutrino scattering based on
the RS model is in a very important point different from the
electron scattering analysis based on the experimental data. The
aim of the RS model is to describe  SPP channels only. In the
electron scattering studies the experimental data is that of the
inclusive cross section. In order to perform the analogous
analysis for the RS model it is necessary to extract and add
contributions from more inelastic channels in the kinematical
region of invariant hadronic mass up to $2~\mathrm{GeV}$.

Our analysis is based on two basic assumptions. The first is that
the RS model predictions for SPP cross sections are fairly close to
what will come out from future precise cross section measurements.
The second is that we know the probability that at a given point in
the kinematically allowed region the final state is that of SPP.
Throughout this paper we will call the probabilities {\it 1-pion
functions}. We obtained these functions numerically \cite{1PF} using
the LUND algorithm \cite{Fun}. The functions (one for each exclusive
SPP channel) were discussed also in \cite{Sar} and they turn out to
depend only on the invariant hadronic mass. For the value of
invariant mass $W\sim 1.6~\mathrm{GeV}$ the contribution from more
inelastic channels amounts to about 50\%. Using these functions one
can rescale SPP contributions to the structure functions and
evaluate the overall structure functions in the region $W\leq 2$
GeV. In section \ref{Single pion production and DIS} we discuss the
precision with which the 1-pion functions are reconstructed. We
present the experimental data for the proton photoproduction 1-pion
function \cite{Photo}. We present also an alternative approach to
calculate the 1-pion function for available hadron multiplicity data
and KNO model \cite{Mul}.

The aim of our paper is to investigate the question if the duality
holds also in $\nu$N reactions. Our methodology is to repeat the
analysis done for eN scattering. We calculate structure functions as
they are defined by the RS model and compare with the DIS structure
functions based on GRV94 PDFs and evaluated at
$Q_{DIS}^2=10$~$\GeV^2$. We perform also a quantitative comparison
of integrated strengths over resonances. All the comparisons are
done first for proton and neutron structure function and then for
their average i.e. for isoscalar (deuterium-like) target. Both
charged current (CC) and neutral current (NC) reactions are
discussed.

It is well known that CC SPP channels on proton and neutron have
distinct properties. The strength od $\Delta$ resonance for proton
reaction is three times as big as its neutron reaction counterpart
due to isospin rules. In the neutrino-proton reaction there is no
need to introduce non-resonant background which gives significant
contribution to the neutrino-neutron SPP channels. The way in
which non-resonant background is treated in the RS model in not
completely satisfactory. For this reason we address the idea of
two-component duality proposed by Harari and Freund \cite{HF}.
They suggested that resonance and non-resonance contributions to
the low energy $\pi$N scattering amplitude (s-channel) correspond
to contributions given by the high energy amplitudes (t-channel)
due to Reggeon and Pomeron exchange respectively. Using the modern
language it can be expressed as existence of a relation between
resonance/valence quark and non-resonant/sea quark contributions
to the structure functions. A confirmation of this idea in eN
interactions was found in \cite{Nic2}: the $F_2$ structure
function averaged over resonances at low values of the Nachtmann
variable ($\xi \leq 0.3$) behaves in the way which strongly
resembles the behavior of valence quark contribution to DIS
scaling curve. There is also a striking similarity between the
above mentioned averaged $F_2$ rescaled by a factor of
$\frac{18}{5}$ and the $xF_3$ $\nu$N data. If resonance
contribution to structure function is dual to the valence DIS
contribution and if the overall duality is satisfied then also
non-resonant background should be dual to the sea quark
contribution. The last duality could be then used to provide a
model for non-resonant background.

%It is an attractive idea to use the sea DIS contribution to
%describe the non-resonant background seen in SPP channels
%\cite{WRONG}.

The question of the QH duality in $\nu$N interactions was
discussed already by Matsui, Sato and Lee \cite{LS} and by
Lalakulich, Paschos and Piranishvili \cite{LP}. In \cite{LS} the
Lee-Sato model \cite{LS2} for $\Delta$ production in eN and $\nu$N
scattering was analyzed. It was shown that in the vicinity of
$\Delta$ excitation peak the local duality holds for both proton
and iso-scalar structure functions in CC and NC neutrino
reactions. The resonance model investigated in \cite{LP} includes
four resonances $P_{33}$, $P_{11}$, $D_{13}$ and $S_{11}$ of
$W<1.6$ GeV. The model (unlike the RS model) contains correction
for the non-zero charged lepton mass. The conclusions are in
agreement with those contained in \cite{LS}. Some qualitative
elements of the present analysis of the RS model can be found in
\cite{Lip}. In \cite{Bodek-RS} several theoretical and practical
issues related to the problem of how to combine smoothly the RS
and DIS contributions in Monte Carlo generators are addressed.
Using the idea of duality one can combine experimental and
theoretical arguments and find suitable modification of the
structure functions which average over resonances for small values
of $Q^2$. One expects that duality should hold for $Q^2\geq 0.5$
GeV$^2$ \cite{CI}. For $Q^2$ approaching zero $F_2$ structure
function for electro-production should behave like $F_2\sim Q^2$
due to gauge invariance. The presence of axial current modifies
this behavior. An important piece of information is provided by
the Adler sum rule \cite{Adler}. It is argued  that vector and
axial parts of the structure functions should be modified in a
different way \cite{Bodek-RS}.

Our paper is organized as follows. In Section 2 the necessary
theoretical introduction is given. The recipe to obtain structure
functions from the Rein-Sehgal model is presented and the 1-pion
functions are introduced. The functions $\mathcal{R}_{2,3}$ which
measure how well the duality holds are also defined and ambiguities
in their definition are discussed. In Section 3 we present the
results of the numerical analysis and their discussion. Section 4
contains the conclusions.

\section{Theoretical foundations of the numerical analysis}
\label{Theoretical foundations of the numerical analysis}
\subsection{Rein-Sehgal structure functions}
\label{Rein-Sehgal structure functions}
%%%%%%%%%%%%%%%%%%%%%%%%%%%%%%%%%%%%%%%%%%%%%%%%%%%%%%%%%%%%%%%%%%%%%%%%%%%%%%
\begin{figure}
\center
\includegraphics{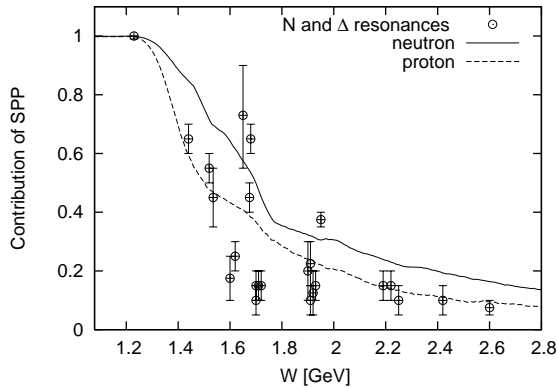}%2
\caption{Elasticities of resonances \cite{PDG} and the 1-pion
functions \cite{1PF} for neutrino CC reactions.
\label{elastycznosc}}
\end{figure}
%%%%%%%%%%%%%%%%%%%%%%%%%%%%%%%%%%%%%%%%%%%%%%%%%%%%%%%%%%%%%%%%%%%%%%%%%%%%%%%

%%%%%%%%%%%%%%%%%%%%%%%%%%%%%%%%%%%%%%%%%%%%%%%%%%%%%%%%%%%%%%%%%%%%%%%%%
\begin{figure}
\includegraphics{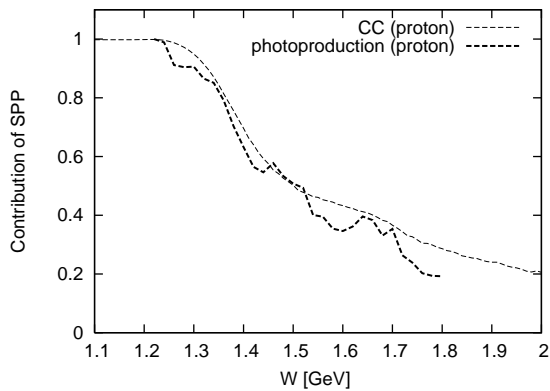} %{res_proton_elasticity.eps}
\caption{ Comparison of the 1-pion functions for neutrino-proton CC
scattering (solid line)  and proton photoproduction (dashed line).
The 1-pion function for the photoproduction was extracted based on
the data from \cite{Photo}. \label{1pion_photoproduction}}
\end{figure}
%%%%%%%%%%%%%%%%%%%%%%%%%%%%%%%%%%%%%%%%%%%%%%%%%%%%%%%%%%%%%%%%%%%%%%%%%

%%%%%%%%%%%%%%%%%%%%%%%%%%%%%%%%%%%%%%%%%%%%%%%%%%%%%%%%%%%%%%%%%%%%%%%%%%%%%
\begin{figure}
\includegraphics{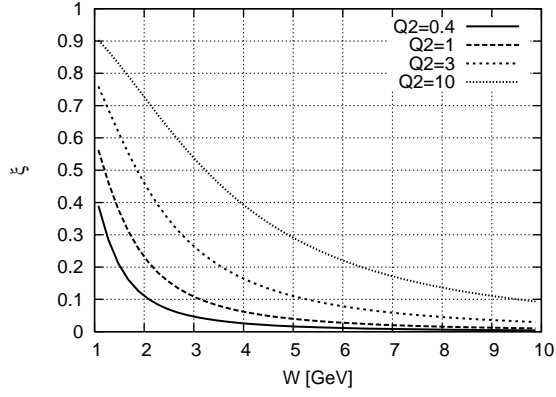}%2
\caption{Dependence of the Nachtmann variable $\xi$ on hadronic
invariant mass calculated at $Q^2$ = 0.4, 1, 3 and 10 $\GeV^2$.
\label{area}}
\end{figure}
%%%%%%%%%%%%%%%%%%%%%%%%%%%%%%%%%%%%%%%%%%%%%%%%%%%%%%%%%%%%%%%%%%%%%%%%%%%%

We consider the following SPP charged current and neutral current
reactions:
\begin{eqnarray}
\nu (k) + \mathcal{N}(p) \to l(k') + \mathcal{N}^* (p')\to l +
\mathcal{N'} + \pi
\nonumber \\
\nu (k) + \mathcal{N}(p) \to \nu (k') + \mathcal{N}^* (p')  \to
\nu + \mathcal{N}' + \pi \nonumber
\end{eqnarray}
In the LAB frame the momentum transfer is:
\begin{equation}
q^\mu=k^{\mu}-k'^{\mu}=(\nu,0,0, q), \quad q_{\mu}q^{\mu} = \nu^2
- q^2\equiv - Q^2.
\end{equation}
The leptonic current is defined as:
\begin{equation}
{\cal J}^{\mu}_{lepton}=\bar{u}(k')\gamma^{\mu}(1-\gamma_5)u(k).
\end{equation}
In the RS model the leptonic mass is set to be zero. In this limit
\begin{equation}
q_{\mu}{\cal J}^{\mu}_{lepton}=0.
\end{equation}
One can introduce the basis of three vectors of length $\pm 1$
orthogonal to $q^{\mu}$:
\begin{eqnarray*}
e_L^{\mu}&=&\frac{1}{\sqrt{2}}(0,1,-i,0), \\
e_R^{\mu}&=&\frac{1}{\sqrt{2}}(0,-1,-i,0), \\
e_S^{\mu}&=&\frac{1}{\sqrt{Q^2}}(q,0,0,\nu).
\end{eqnarray*}
Correspondingly, the leptonic tensor can be decomposed as:
\begin{equation}
L^{\mu\nu}=k^{\mu}k'^{\nu}+k'^{\mu}k^{\nu}-g^{\mu\nu}k\cdot
k'-i\varepsilon^{\mu\nu\kappa\lambda}k_{\kappa}k'_{\lambda}=
\end{equation}
\begin{equation}
=\sum_{\alpha ,\beta\in (S,L,R)} M^{\alpha\beta
}e_\alpha^{\mu}(e_\beta^{\nu})^*.
\end{equation}
When we calculate the contraction of the leptonic tensor with the
hadronic tensor
\begin{equation} W_{\mu\nu} = \left(-g_{\mu\nu}W_1 + \frac{p_\mu
p_\nu}{M^2}W_2 \ - \frac{\mathrm{i}\epsilon_{\mu\nu \alpha
\beta}p^\alpha q^\beta}{2M^2} W_3\right),
\end{equation}
($M$ is the nucleon mass) we find that
\begin{equation}
L^{\mu\nu}W_{\mu\nu}=L^{\mu\nu}_{diag}W_{\mu\nu},
\end{equation}
where
\begin{equation}
L^{\mu\nu}_{diag}=A^2e_S^{\mu}(e_S^{\nu})^*+B^2e_L^{\mu}(e_L^{\nu})^*+C^2e_R^{\mu}(e_R^{\nu})^*.
\end{equation}

$A^2$, $B^2$, $C^2$ are Lorentz scalars which can be evaluated in
the LAB frame:
\begin{eqnarray}A^2 &=& L_{\mu\nu}e_S^{\mu}
(e_S^{\nu})^*= \frac{Q^2}{2q^2}\left( (2E-\nu)^2-q^2\right),\\
B^2 &=& L_{\mu\nu}e_L^{\mu} (e_L^{\nu})^*= \frac{Q^2}{4q^2}(
2E-\nu+q)^2,\\
C^2 &=& L_{\mu\nu}e_R^{\mu} (e_R^{\nu})^*= \frac{Q^2}{4q^2}(
2E-\nu-q)^2.
\end{eqnarray}
Decomposition of the leptonic tensor entails the decomposition of
the cross section into three contributions $\sigma_L, \sigma_R,
\sigma_S$ which are interpreted as cross sections of intermediate
boson in given polarization states scattered off nucleon. In its
final form the Rein-Sehgal formula for the cross section reads:

\begin{equation}
d^2 \sigma = \frac{G_F^2 \cos^2 \theta_C}{4 \pi^2}
\left(\frac{Q^2}{q^2} \right) \kappa\left(u^2 \sigma_L + v^2
\sigma_R + 2uv\sigma_S \right)dQ^2d\nu,
\end{equation}
where
$$
u^2=B^2\frac{q^2}{E^2Q^2}, \quad v^2=C^2\frac{q^2}{E^2Q^2},\quad
uv=A^2\frac{q^2}{E^2Q^2}, \quad \kappa=\nu -\frac{Q^2}{2M} .
$$
Thus we can write:
\begin{equation}
d^2 \sigma = \frac{G_F^2 \cos^2 \theta_C}{4 \pi^2 E^2}\kappa
 \left(B^2 \sigma_L + C^2 \sigma_R +
A^2\sigma_S \right)dQ^2d\nu.
\end{equation}
$\sigma_{L,R,S}$ are then calculated within the quark model and
are given in the explicit way for each SPP channel separately.

In order to identify $\sigma_{L,R,S}$ as linear combinations of
the structure functions we calculate:

\begin{eqnarray*}
L_{\mu\nu}W^{\mu\nu}=A^2\left(W_2\frac{q^2}{Q^2}-W_1\right) +
B^2\left(W_1+W_3\frac{q}{2M}\right)+C^2\left(W_1-W_3\frac{q}{2M}\right)
\end{eqnarray*}
and
\begin{eqnarray}
\frac{d^2 \sigma}{d\nu dQ^2} &=& \frac{G_F^2 \cos^2 \theta_C}{4 \pi
E^2} \left[ A^2\left(W_2\frac{q^2}{Q^2}-W_1\right)
+B^2\left(W_1+W_3\frac{q}{2M}\right)
+C^2\left(W_1-W_3\frac{q}{2M}\right)\right].
\end{eqnarray}

A simple comparison gives the following identification of the
$\sigma's$ in terms of $W's$:
\begin{eqnarray}
-W_1+W_2\frac{q^2}{Q^2}&=&\frac{\kappa}{\pi} \sigma_S,\\
W_1+W_3\frac{q}{2M}&=& \frac{\kappa}{\pi}  \sigma_L,\\
W_1-W_3\frac{q}{2M}&=&\frac{\kappa}{\pi}  \sigma_R.
\end{eqnarray}
Using the definition of the structure functions $F's$ we write down
the final expressions for the  RS model structure functions:
\begin{eqnarray}
F_1^{RS}& = & MW_1=M \frac{\kappa}{2\pi} (\sigma_L + \sigma_R),\\
F_2^{RS}& = & \nu W_2=\nu  \frac{\kappa}{2\pi}\frac{Q^2}{q^2}
\left(2
\sigma_S + \sigma_L + \sigma_R \right),\\
F_3^{RS}&=& \nu W_3=\nu  \frac{\kappa}{\pi} \frac{ M}{q}(\sigma_L
- \sigma_R).
\end{eqnarray}

\subsection{DIS structure functions}
\label{DIS structure functions}

We apply the simple model for the DIS structure functions: they
are given by appropriate linear combinations of the parton
distribution functions (PDFs). In the kinematical region we are
interested in the charm contribution can be neglected. For CC
$\nu$N interaction the DIS structure functions are \cite{Leader}:
\begin{eqnarray}
F_2^{CC} (\nu p) & = & 2x\left(d\cos^2\theta_c + s\sin^2\theta_c + \bar{u} \right),\\
\label{xF3p} x F_3^{CC} (\nu p)& = & 2x\left(d\cos^2\theta_c +
s\sin^2\theta_c -
\bar{u} \right), \\
F_2^{CC}(\nu n)   & = & 2x \left( u\cos^2\theta_c + s\sin^2\theta_c + \bar{d} \right), \\
\label{xF3n} x F_3^{CC}(\nu n) & = & 2 x \left( u\cos^2\theta_c +
s\sin^2\theta_c - \bar{d} \right).
\end{eqnarray}

\noindent For $\nu$N NC interaction the DIS structure functions
are:

\begin{eqnarray}
F_2^{NC}(\nu p)   & = & 2x \left( (g_L^2 + g_R^2)\left(u + \bar{u}
\right) + ({g'}_L^2 +
{g'}_R^2)\left(d +  \bar{d} + 2s \right)\right), \\
x F_3^{NC}(\nu p) & = & 2 x  \left( (g_L^2 - g_R^2)\left(u -
\bar{u} \right) + ({g'}_L^2 - {g'}_R^2)\left(d  -
\bar{d} \right)\right),  \\
F_2^{NC}(\nu n)   & = &  2x \left( (g_L^2 + g_R^2)\left(d +
\bar{d} \right) + ({g'}_L^2 + {g'}_R^2)\left(u +
\bar{u} + 2s\right)\right),  \\
x F_3^{NC}(\nu n ) & = & 2  x \left((g_L^2 - g_R^2)\left(d -
\bar{d} \right) + ({g'}_L^2 - {g'}_R^2)\left(u  - \bar{u}
\right)\right).
\end{eqnarray}
with
\begin{eqnarray}
g_L  =  \frac{1}{2} - \frac{2}{3}\sin^2\theta_W,  & & g_R  =   - \frac{2}{3}\sin^2\theta_W, \\
{g'}_L  =   -\frac{1}{2} + \frac{1}{3}\sin^2\theta_W, & & {g'}_R =
\frac{1}{3}\sin^2\theta_W.
\end{eqnarray}
Wherever we present the plots of $F_1$ we use the Callan-Gross
relation:
\begin{equation}
F_2 = 2 x F_1.
\end{equation}
In the quantitative analysis we restrict ourselves to $F_2$ and
$xF_3$ only.

We use GRV94 (LO) PDF's~\cite{GRV94} which are defined for
$Q^2>0.23$ GeV$^2$ and $x \geq 10^{-5}$ and distinguish valence
and sea quark contributions. GRV94 PDF's are used in many Monte
Carlo generators of events \cite{NuInt}.

\subsection{1-pion functions}
\label{Single pion production and DIS}

The 1-pion functions are defined for each SPP channel separately
as probabilities that at a given value of $W$ the final hadronic
state is that of SPP:
\begin{equation}
\label{eq_1pion} f_{1\pi}(W)\equiv\frac {\displaystyle \frac{d
\sigma^{SPP}}{dW}}{\displaystyle\frac{d\sigma^{DIS}}{dW}}.
\end{equation}

The 1-pion functions used in this paper were obtained from the Monte
Carlo simulation based on the LUND algorithm. Therefore they are
defined by fragmentation and hadronization routines implemented
there. The comparison of the 1-pion functions with elasticity
factors of resonances included in the RS model is shown in Fig.
\ref{elastycznosc}. The agreement is satisfactory. It was also
checked that  the simulations based on LUND give rise to charged
hadron multiplicities consistent with the experimental data
\cite{GJS}.

We assume that the following relation holds between the structure
functions of the RS model and the overall structure functions
$F_{j}^{RES}$ for the inclusive cross section:
\begin{equation}\label{renorm}
F_{j}^{RES}(x,Q^2) = \frac{\displaystyle
F^{RS}_{j}(x,Q^2)}{\displaystyle f_{1\pi}(W(x,Q^2))},
\end{equation}
where $j=1,2,3$. In the case of neutron (CC reactions) and NC
structure functions in the above formula we apply the sum of the
1-pion functions for two SPP channels.

We will see that our results are sensitive to the details of the
1-pion functions in the region of $W\in (1.5,\ 2)$~GeV where the
rescaling effects are most important. We investigated this point in
more detail:

\begin{itemize}

\item[i)] We extracted the 1-pion function from $\gamma p$
photo-production data \cite{Photo}, see Fig.
\ref{1pion_photoproduction}. We conclude that it is rather similar
to the function we used in our numerical computations.

\item[ii)] We tried to evaluate the 1-pion functions from
available hadron multiplicities data in neutrino reactions
\cite{Mul} extrapolating the predictions to the region $W\in (2,\
3)$~GeV.

We know the average charged hadron multiplicities

\[\langle n_{ch}\rangle_{\nu p}=-0.05\pm 0.11 + (1.43\pm 0.04)\ln
(W^2),\]

\[\langle n_{ch}\rangle_{\nu n}=-0.2\pm 0.07 + (1.42\pm 0.03)\ln
(W^2),\]

and the neutral pion multiplicity measured in $\nu p$ reactions:

\[\langle n_{\pi^0}\rangle_{\nu p}=0.14\pm 0.26 + (0.5\pm 0.08)\ln
(W^2).\]

We assume KNO distribution of multiplicities and  that in the first
approximation there are only nucleons and pions in the final state.
For $\nu p$ reaction it is also necessary to make some assumptions
about charged pions. We expect that the fraction of charged pions is
the same as in the $\mu p$ reaction for which the experimental data
is available. We obtained the following values for the 1-pion
functions at $W=2$~GeV: $0.14$ for $\nu p$ and $0.38$ for $\nu n$.
The obtained values must be reduced by $\sim 10~\%$ due to the
presence of other exclusive channels. We conclude that the KNO
results seems to be in general agreement with basic properties of
our 1-pion functions: the fraction of SPP channels on neutron is
bigger then on the proton SPP  and the orders of magnitude are quite
similar.

\end{itemize}

\subsection{Kinematics}
\label{Kinematics}

In the QH duality analysis different kinematical regions are
simultaneously involved in the discussion.

A common presentation of the duality is done by means of the
comparison of plots of structure functions in the Nachtmann
variable:
\begin{equation}
\xi(x,Q^2) = \frac{2x}{1 + \sqrt{1 + 4 x^2 M^2/ Q^2}}
\end{equation}
which takes into account target mass corrections.

The resonance region is defined in terms of invariant hadronic
mass as $W\in (M+m_{\pi}, 2\, \GeV$) which is the natural choice
for the Rein-Sehgal model. Other options ($W_{max}<2\,\GeV)$ for
the definition of the resonance region were considered in
\cite{GJS}.

Fig. \ref{area} illustrates the dependence of the Nachtmann
variable on hadronic invariant mass at fixed values of $Q^2$. For
typical $Q^2$ for the resonance region $Q^2\in (0.5, 3)\,\GeV^2$
one obtains the following domain in the Nachtmann variable:
$\xi\in (0.13, 0.76)$. This region in $\xi$ when combined with
$Q^2_{DIS}=5,\ 10,\ 20$~GeV$^2$ corresponds to:
\begin{eqnarray}
Q^2&=&5\,\GeV^2\quad \ \,\;\; \longrightarrow  \quad W\in (1.3,\, 5.8)\,\GeV, \\
Q^2&=&10\,\GeV^2\quad\ \ \longrightarrow \quad W\in (1.8,\,
8.2)\,\GeV, \\
Q^2&=&20\,\GeV^2\quad \ \ \longrightarrow  \quad W\in (2.5,\,
11.6)\,\GeV .
\end{eqnarray}

In our numerical analysis we use $Q^2_{DIS}=10\,\GeV^2$.
%%%%%%%%%%%%%%%%%%%%%%%%%%%%%%%%%%%%%%%%%%%%%%%%%%%%%%%%%%%%%%%%%%%%%%%%%%%%%%%%%%%
\begin{figure}
\includegraphics{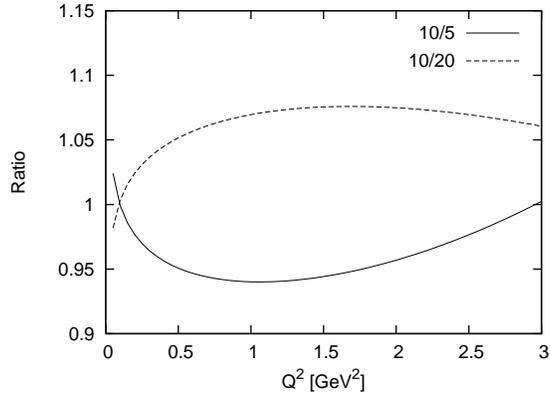}%2
\caption{Uncertainties in (\ref{ratio_2}) due to different
definitions of $Q^2_{DIS}$. Solid line corresponds to (\ref{10/5})
and dashed line to (\ref{10/20}). \label{Ratio_test}}
\end{figure}
%%%%%%%%%%%%%%%%%%%%%%%%%%%%%%%%%%%%%%%%%%%%%%%%%%%%%%%%%%%%%%%%%%%%%%%%%%%%%%%%%%%%

\subsection{Quark-Hadron Duality}
\label{Quark-Hadron Duality}

The QH duality is said to be present on the quantitative level if
the following relation between resonance and scaling structure
functions holds:
%%%%%%%%%%%%%%%%%%%%%%%%%%%%%%%%%%%%%%%%%%%%%%%%%%%%%%%%%%%%%%%%%%%%%%%%%%%%%%%%%%%%
\begin{figure*}
\includegraphics{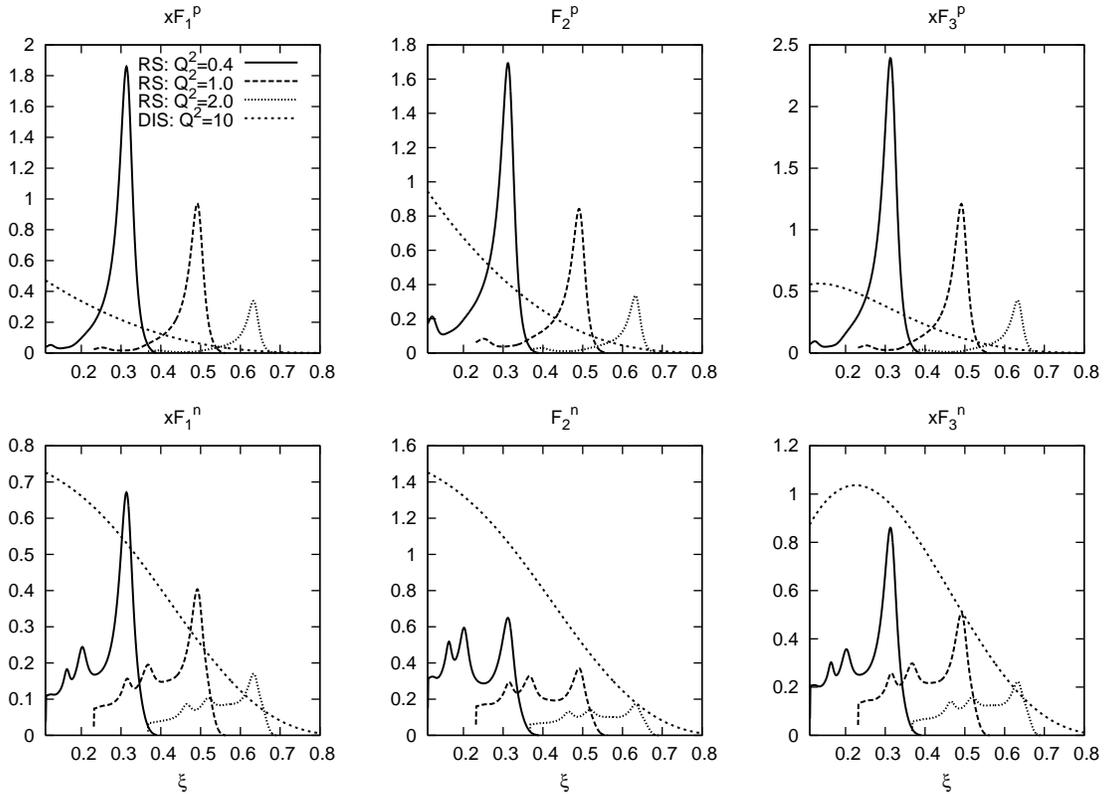}
\caption{Comparison of the Rein-Sehgal structure functions at
$Q^2=$ 0.4, 1 and 2~GeV$^2$  with the appropriate scaling
functions at $Q^2_{DIS}$=10~GeV$^2$. In the first row $xF_1$,
$F_2$ and $xF_3$ structure functions for CC neutrino-proton
scattering are plotted. In the second row the structure functions
for CC neutrino-neutron scattering are shown.
\label{Slizganie_CC}}
\end{figure*}
%%%%%%%%%%%%%%%%%%%%%%%%%%%%%%%%%%%%%%%%%%%%%%%%%%%%%%%%%%%%%%%%%%%%%%%%%%%%%%%%%%%%
\begin{equation}
\label{FESR} \int_{\xi_{min}}^{\xi_{max}} d \xi F_i^{RES}(\xi ,
Q^2_{RES}) \approx \int_{\xi_{min}}^{\xi_{max}} d \xi F_i^{DIS}(\xi,
Q^2_{DIS}).
\end{equation}
The above equation should hold for different values of $Q^2_{RES}$
characteristic for the resonance production and for a fixed value of
$Q^2_{DIS}$. The region of integration -- RES region -- is defined
to be identical with the resonance region of the RS model:
$W_{min}=M+m_{\pi}$ and $W_{max}=2~\mathrm{GeV}$, which is then
translated into appropriate region in $\xi$:
\begin{eqnarray}
\label{def_area} \xi_{min}=\xi(W_{max}, Q^2_{RES}), \quad
\xi_{max}=\xi(W_{min}, Q^2_{RES})
\end{eqnarray}
In the quantitative analysis we define ratios of two integrals
over the resonance region:
\begin{equation}
\label{ratio_def} \mathcal{R}\left(f,Q^2_{R};
g,Q^2_{D}\right)=\frac{\displaystyle \displaystyle
\int_{\xi_{min}}^{\xi_{max}} d \xi\, f(\xi, Q^2_{R})
}{\displaystyle\int_{\xi_{min}}^{\xi_{max}} d \xi\, g (\xi,
Q^2_{D})}.
\end{equation}
Using the above quantity, we define the function:
\begin{eqnarray}
\mathcal{R}_2(Q^2_{RES}, Q^2_{DIS})& \equiv
&\mathcal{R}\left(F^{RES}_2,Q^2_{RES};
F_2^{DIS},Q^2_{DIS}\right). \nonumber \\
\label{ratio_2}
\end{eqnarray}

and

\begin{eqnarray}
\mathcal{R}_3(Q^2_{RES}, Q^2_{DIS})& \equiv &\mathcal{R}\left(x
F^{RES}_{3},Q^2_{RES}; x
F_3^{DIS},Q^2_{DIS}\right).\nonumber\\
\label{ratio_3}
\end{eqnarray}

There is an ambiguity in the quantitative analysis of duality
because the above functions depend on the arbitrarily chosen value
of $Q^2_{DIS}$. The differences between scaling curves calculated
at different $Q^2$ are not relevant for making qualitative
statements about the duality but do matter in quantitative
analysis. To illustrate this we calculate
\begin{equation}
\label{10/5} \mathcal{R}_{10/5} \equiv
\mathcal{R}\left(F_2^{DIS},Q^2_{DIS}=10;
F_2^{DIS},Q^2_{DIS}=5\right),
\end{equation}
and
\begin{equation}
\label{10/20} \mathcal{R}_{10/20} \equiv
\mathcal{R}\left(F_2^{DIS},Q^2_{DIS}=10;
F_2^{DIS},Q^2_{DIS}=20\right),
\end{equation}
the integration region of the above integrals is defined by
$Q^2_{RES}$. The results are shown in Fig. \ref{Ratio_test}. We
see that the ambiguity is of the order of $7\%$ and is largest for
$Q^2_{RES}\sim 1.5$ GeV$^2$. This limits the precision of the
quantitative statements about the QH duality.

%%%%%%%%%%%%%%%%%%%%%%%%%%%%%%%%%%%%%%%%%%%%%%%%%%%%%%%%%%%%%%%%%%%%%%%%%%%%%%%%%%
\begin{figure}
\includegraphics{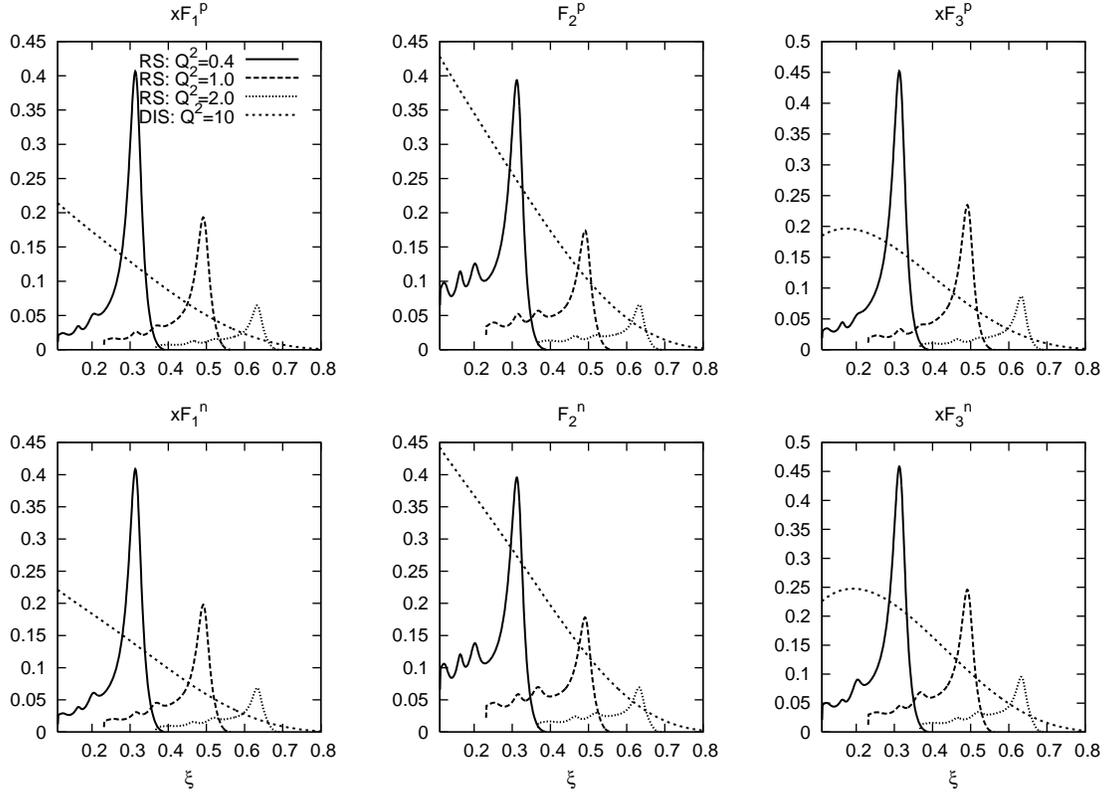}%2
\caption{The same as in Fig.~\ref{Slizganie_CC} but for NC
reactions.\label{Slizganie_NC}}
\end{figure}
%%%%%%%%%%%%%%%%%%%%%%%%%%%%%%%%%%%%%%%%%%%%%%%%%%%%%%%%%%%%%%%%%%%%%%%%%%%%%%%%%%%

In the investigation of two-component duality we single out
resonant and non-resonant contributions to the RS model structure
functions
\begin{equation}
F_j^{RES}=F_{j,res}+F_{j,nonres}
\end{equation}
and we also separate valence and sea quark contributions to the
DIS structure functions:
\begin{equation}
F_j^{DIS}=F_{j,sea}+F_{j,val}.
\end{equation}

%%%%%%%%%%%%%%%%%%%%%%%%%%%%%%%%%%%%%%%%%%%%%%%%%%%%%%%%%%%%%%%%%%%%%%%%%
\begin{figure*}
\centerline{
\includegraphics{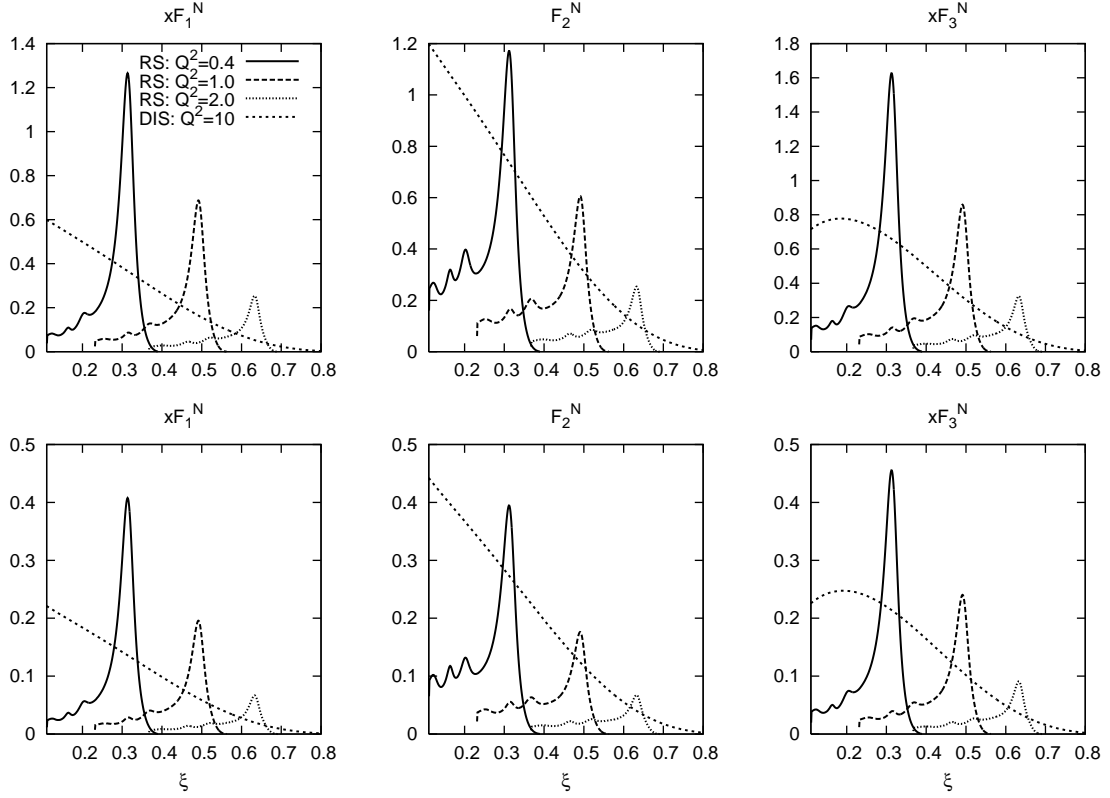}%3
}\caption{ Comparison of the Rein-Sehgal structure functions at
$Q^2=$ 0.4, 1 and 2~GeV$^2$ with the appropriate scaling functions
at $Q^2_{DIS}$=10~GeV$^2$. In the first row the plots of the
$xF_1$, $F_2$ and $xF_3$ structure functions for CC
neutrino-isoscalar target scattering are presented. In the second
row structure functions for NC neutrino-isoscalar target
scattering are shown. \label{slizganie_izo}}
\end{figure*}
%%%%%%%%%%%%%%%%%%%%%%%%%%%%%%%%%%%%%%%%%%%%%%%%%%%%%%%%%%%%%%%%%%%%%%%%

We calculate the following functions:
\begin{equation}
\label{eq_ratio_val_F2} \mathcal{R}_{2}^{val}(Q^2_{RES},
Q^2_{DIS})\equiv\mathcal{R}\left(F_{2,res},Q^2_{RES};
F_{2,val},Q^2_{DIS}\right).
\end{equation}
and
\begin{equation}
\label{eq_ratio_val_F3} \mathcal{R}_{3}^{val}(Q^2_{RES},
Q^2_{DIS})\equiv\mathcal{R}\left(xF_{3,res},Q^2_{RES};
xF_{3,val},Q^2_{DIS}\right).
\end{equation}

%%%%%%%%%%%%%%%%%%%%%%%%%%%%%%%%%%%%%%%%%%%%%%%%%%%%%%%%%%%%%%%%%%%%%%%%%

\begin{figure}
\centerline{
\includegraphics{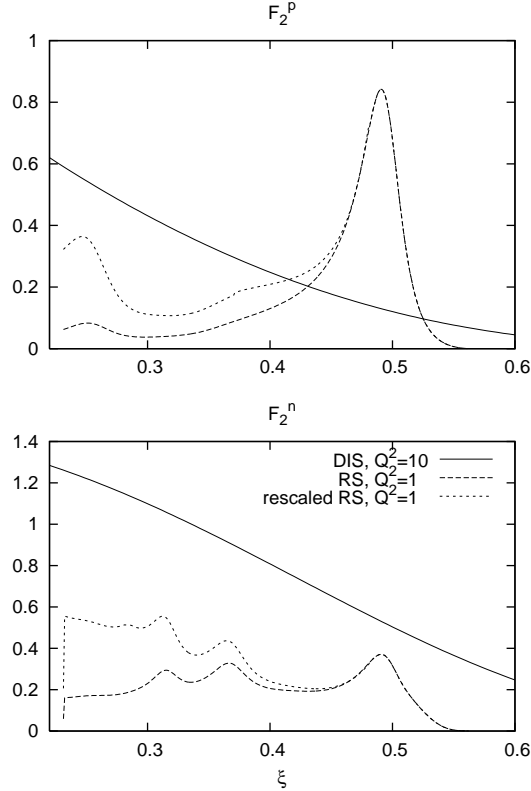}%3
}\caption{The comparison of the proton and neutron CC $F_2$
structure functions for the RS model (dashed line), rescaled RS
model (dotted line) and the DIS (solid line). The RS structure
functions are calculated at $Q^2=1\,\GeV^2$ while the DIS curves
are plotted for $Q^2=10\,\GeV^2$. \label{1pion_example}}
\end{figure}

\section{Numerical results and discussion}
\label{Numerical results and discussion}

In the numerical analysis we confine ourselves to the case of
neutrino interactions and leave out the antineutrino ones.

In Figs.~\ref{Slizganie_CC} -- \ref{slizganie_izo} we present a
comparison of the scaling structure function with the RS structure
functions calculated at $Q^2_{RES}=0.4,~ 1~ \mathrm{and}~
2$~GeV$^2$. The Figs. \ref{Slizganie_CC} and \ref{Slizganie_NC}
correspond to CC and NC reactions respectively with proton
structure functions in the upper row and neutron structure
functions below.

%%%%%%%%%%%%%%%%%%%%%%%%%%%%%%%%%%%%%%%%%%%%%%%%%%%%%%%%%%%%%%%%%%%%%%%%%%%%%%
\begin{figure}
\includegraphics{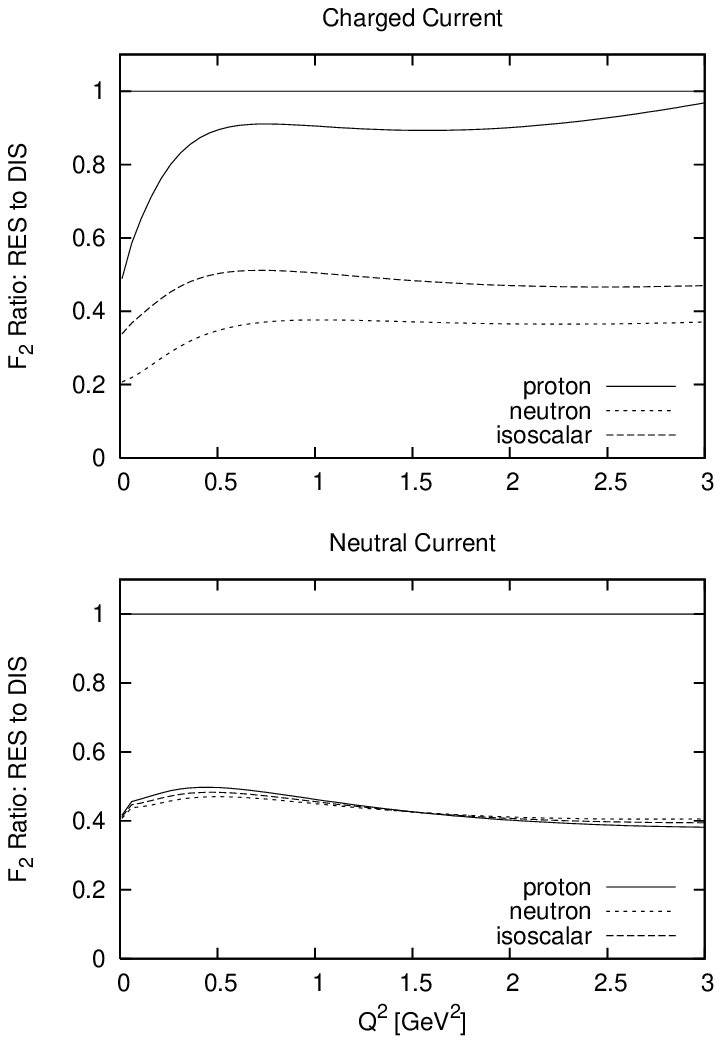}%2
\caption{The functions $\mathcal{R}_{2}$  for different targets and
reactions. The ratios are calculated for CC and NC structure
functions in the cases of proton (solid lines), neutron (dotted
lines) and isoscalar target (dashed lines). \label{ratio_F2}}
\end{figure}
%%%%%%%%%%%%%%%%%%%%%%%%%%%%%%%%%%%%%%%%%%%%%%%%%%%%%%%%%%%%%%%%%%%%%%%%%%%%%%%

%%%%%%%%%%%%%%%%%%%%%%%%%%%%%%%%%%%%%%%%%%%%%%%%%%%%%%%%%%%%%%%%%%%%%%%%%%%%%%
\begin{figure}
\includegraphics{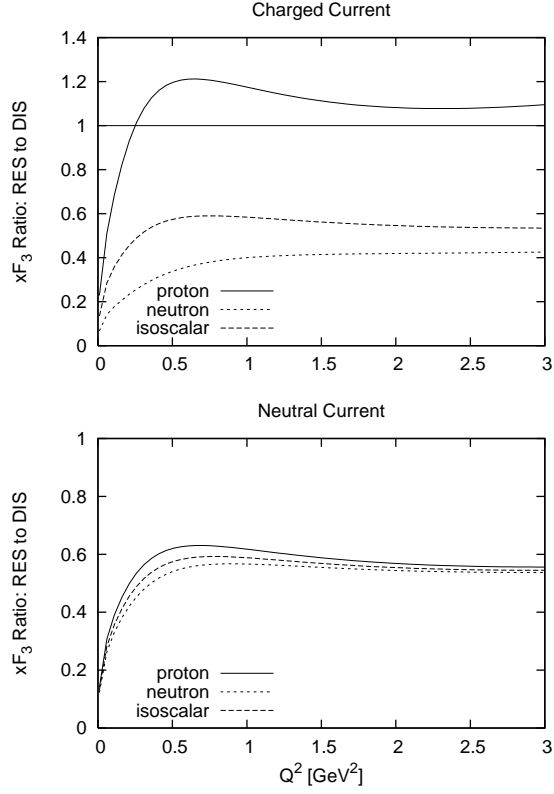}%2
\caption{The same as in Fig. \ref{ratio_F2} but for $xF_3$ (ratio
$\mathcal{R}_{3}$). \label{ratio_F3}}
\end{figure}
%%%%%%%%%%%%%%%%%%%%%%%%%%%%%%%%%%%%%%%%%%%%%%%%%%%%%%%%%%%%%%%%%%%%%%%%%%%%%%%%%%%

%%%%%%%%%%%%%%%%%%%%%%%%%%%%%%%%%%%%%%%%%%%%%%%%%%%%%%%%%%%%%%%%%%%%%%%%%%%%%%%
\begin{figure}
\includegraphics{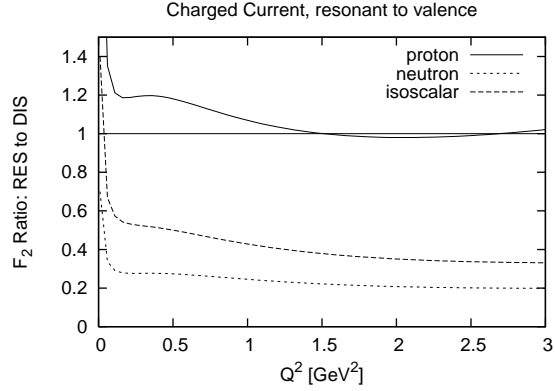}%2
\caption{The plots of functions $\mathcal{R}_{2}^{val}$ defined in
Eq. \ref{eq_ratio_val_F2}. The computations are performed for the
CC reactions for proton (solid line) neutron (dotted line) and
isoscalar targets (dashed line).\label{ratio_val_F2}}
\end{figure}
%%%%%%%%%%%%%%%%%%%%%%%%%%%%%%%%%%%%%%%%%%%%%%%%%%%%%%%%%%%%%%%%%%%%%%%%%%%%%%%%%

%%%%%%%%%%%%%%%%%%%%%%%%%%%%%%%%%%%%%%%%%%%%%%%%%%%%%%%%%%%%%%%%%%%%%%%%%%%%%%
\begin{figure}
\includegraphics{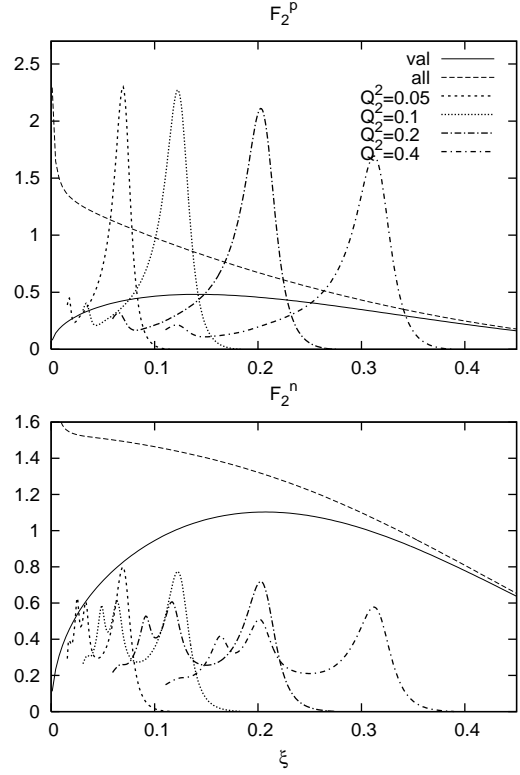}
\caption{Comparison of the proton and neutron CC Rein-Sehgal $F_2$
structure functions at $Q^2=$ 0.05, 0.1, 0.2 and 0.4~GeV$^2$ with
DIS structure (dashed line) at $Q^2_{DIS}$=10~GeV$^2$ and the
contribution valence quarks (solid line).
 \label{Slizganie_F2_CC_double}}
\end{figure}
%%%%%%%%%%%%%%%%%%%%%%%%%%%%%%%%%%%%%%%%%%%%%%%%%%%%%%%%%%%%%%%%%%%%%%%%%%%%%%%%

%%%%%%%%%%%%%%%%%%%%%%%%%%%%%%%%%%%%%%%%%%%%%%%%%%%%%%%%%%%%%%%%%%%%%%%%%%%%%%%
\begin{figure}
\includegraphics{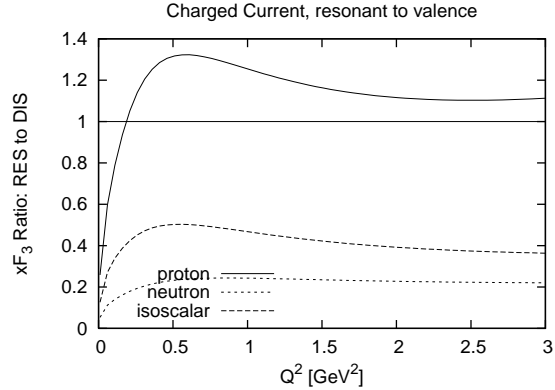}%2
\caption{The same as in Fig. \ref{ratio_val_F2} but for $xF_3$.
\label{ratio_val_F3}}
\end{figure}
%%%%%%%%%%%%%%%%%%%%%%%%%%%%%%%%%%%%%%%%%%%%%%%%%%%%%%%%%%%%%%%%%%%%%%%%%%%%%

%%%%%%%%%%%%%%%%%%%%%%%%%%%%%%%%%%%%%%%%%%%%%%%%%%%%%%%%%%%%%%%%%%%%%%%%%%%%%%%%%
\begin{figure}
\includegraphics{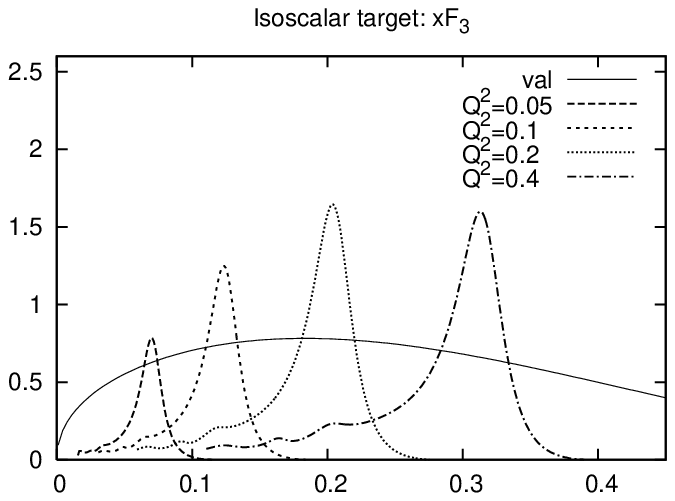}
\caption{Comparison of the isoscalar CC Rein-Sehgal $xF_3$ structure
functions at $Q^2=$ 0.05, 0.1, 0.2 and 0.4~GeV$^2$ with appropriate
valence quark contribution to scaling function (solid line)
calculated at $Q^2_{DIS}=10\,\GeV^2$.
 \label{Slizganie_xF3_CC_double}}
\end{figure}
%%%%%%%%%%%%%%%%%%%%%%%%%%%%%%%%%%%%%%%%%%%%%%%%%%%%%%%%%%%%%%%%%%%%%%%%%%%%%%%

In the case of the RS model for neutrino-proton CC reaction the
$\Delta$ resonance contribution dominates overwhelmingly over
other resonances. One can see the typical manifestation of local
duality: the sliding of the $\Delta$ peaks (calculated at
different $Q^2_{RES}$) along the scaling function.

For neutrino-neutron CC reaction the resonance structure is much
richer. The contributions from the $\Delta$ are usually dominant
but those from more massive resonances are also significant. In
the figure with the $F_2$ structure function three peaks of
comparable size are seen. The DIS contributions dominate over the
RS ones in this case.

Our plots are comparable with those obtained in Ref.~\cite{LS}.
However in our paper the strengths of $\Delta$ peaks for all the
structure functions are proportionally lower. The difference
between structure functions based on CTEQ6 \cite{LS} and GRV94
PDF's is very small.

It is seen that it is virtually impossible to have simultaneously
duality in CC reactions on proton and neutron targets. The strength
of $\Delta$ excitation on proton is approximately three times as big
as for neutron and DIS cross section on neutron is much bigger then
on the proton.

For the NC reactions the dominant $\Delta$ peaks also slide along
scaling function. The structure functions for the proton and
neutron are almost the same.

In Fig.~\ref{slizganie_izo} the analysis is performed for the
isoscalar target. The plots of the structure functions for CC
(upper row) and NC (lower row) reactions are presented. In each
plot the $\Delta$ peak slides along the scaling function and the
local duality is seen.

Apparently (with the exception for CC reaction on proton) there is
little hope for the QH duality in the whole resonance region: the
scaling structure functions are on average larger then the RS ones.
Only the local duality is present after a suitable region in W
around $M_{\Delta}=1.23$ GeV is chosen. But in the Figs.
\ref{Slizganie_CC} -- ~\ref{slizganie_izo} the rescaling of RS
structure functions by means of the 1-pion functions (see Eq.
(\ref{renorm})) is not yet included. The rescaling procedure
increases the RES structure functions making the duality more likely
to appear. We notice also that the rescaling cannot spoil the
statements about the local duality around the $\Delta$ resonance
because the values of the 1-pion function for W in the vicinity of
$M_{\Delta}$ are close to 1 (see Fig. \ref{elastycznosc}).

In Fig.~\ref{1pion_example} we show how the resonance structure
functions are modified by means of the 1-pion functions. The
modifications apply mainly to hadronic invariant mass close to
$2$~GeV.

In order to perform a quantitative analysis of the duality we make
use of the functions $\mathcal{R}_i$ defined in Eqs.
(\ref{ratio_2}-\ref{eq_ratio_val_F3}). We restrict our plots to
the values of $Q^2_{RES}\leq 3\GeV^2$ characteristic for the
resonance production.

In what follows we use the RES structure functions rescaled by means
of the 1-pion functions. We have checked that introduction of the
1-pion function improves the duality significantly. For example for
CC reaction on neutron $\mathcal{R}_{2}$ is increased by a factor of
$\sim\,1.55$ and for proton by $\sim\,1.39$. The difference is
caused by the overwhelming dominance of the $\Delta$ excitation in
the case of proton.

A characteristic feature of most of the plots of $\mathcal{R}_j
(Q^2_{RES})$ is a presence of two qualitatively distinct
behaviors. For $Q^2_{RES}$ smaller then $\sim 0.5$ GeV$^2$ the
functions $\mathcal{R}_j$ vary quickly while for larger values of
$Q^2_{RES}$ they become slowly changing. This seems to correspond
to predictions done in \cite{CI}. Our statements about the duality
will apply only to the region of $Q^2_{RES}\geq 0.5$~GeV$^2$.

In Figs. \ref{ratio_F2} and \ref{ratio_F3} the plots of
$\mathcal{R}_2$ and $\mathcal{R}_3$ for proton, neutron and
isoscalar targets are presented. In the case of CC interaction the
duality is seen on the proton target (accuracy $\leq 20\%$) but
for the neutron and isoscalar targets the duality is absent. In
both cases the average strength of resonance structure functions
amounts to only about a half of the strength of DIS structure
functions. The plots for the NC interactions are almost
independent on the target and in all the cases the DIS
contributions are approximately two times as big as resonance
ones. A different choice of $Q^2_{DIS}$, namely $Q^2_{DIS}=20$
GeV$^2$ makes the values of $\mathcal{R}_{2,3}$ even lower (see
Fig. \ref{Ratio_test}).

The remaining plots address the question of two component duality.
We concentrate on the case of the possible duality between the
resonance and valence quark contributions.

In Fig. \ref{ratio_val_F2} the plot of $\mathcal{R}_2^{val}$ for
the CC interactions is shown. We notice the good duality picture
in the case of proton target but a huge departure from duality in
the case of neutron and isoscalar targets. It is worth noting that
this discrepancy is larger than one shown in Fig. \ref{ratio_F2}
where the general (not two component) notion of duality was
discussed. The novel feature is the apparently singular behavior
at low $Q^2_{RES}$: $\mathcal{R}_2^{val}$ rises quickly in
contrast with $\mathcal{R}_2$ falling down when $Q^2_{RES}$
approaches zero.

The explanation of this follows from the Fig.
\ref{Slizganie_F2_CC_double} where the region of small $Q^2_{RES}$
was analyzed in more detail. We notice that for $Q^2_{RES}$
approaching zero the valence quarks scaling function tends to zero
while the resonance strengths remains virtually unchanged.

Finally in Fig. \ref{ratio_val_F3} the analogous two-component
duality analysis is done for $\mathcal{R}_3^{val}$. The discussion
of $xF_3$ seems to be favorable for the two-component duality
because in the DIS contribution on the isoscalar target there is no
sea quark contribution. We remind also that for the CC reaction on
the proton the non-resonant contribution is absent. In Fig.
\ref{ratio_val_F3} we see that two component duality is satisfied
within $\sim$30\% for the proton target but it is absent for neutron
and isoscalar targets. We notice also that contrary to what we have
seen in the plots for $\mathcal{R}_2^{val}$ now at low $Q^2_{RES}$
all the curves tend to zero.

The explanation of this behavior follows from the Fig.
\ref{Slizganie_xF3_CC_double}. One can see that in the case of
$xF_3$ both the resonance and valence quark structure functions
fall down for $Q^2$ approaching zero. The behavior of $xF_3$ is
the same as that discussed in \cite{Nic2}.

We do not present plots exploring the duality between the
non-resonant part of the resonance model and the sea quark
contribution. No sign of two component duality is seen in this case.
\begin{figure}
\includegraphics[width=9cm, height=10cm]{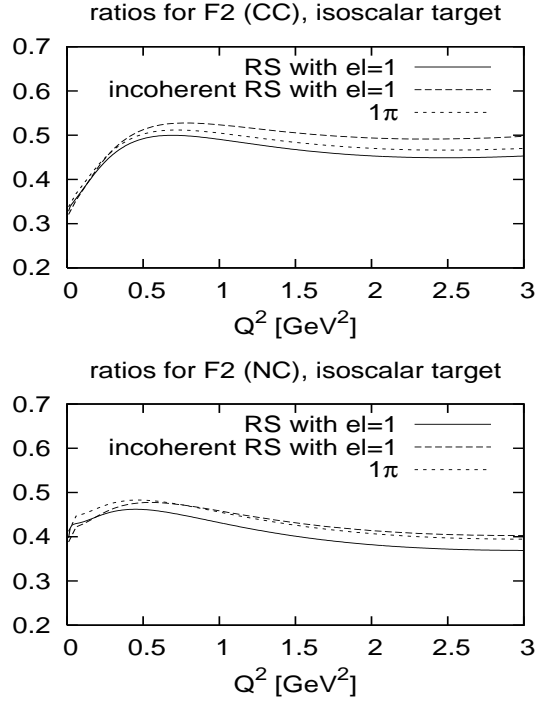}%{ratio_1pion_el_nu_isoscalar.eps}
\caption{The ratios presented in Fig. \ref{ratio_F2} for isoscalar
target are calculated without 1-pion functions but with all the
elasticities equal to 1 with amplitudes added in either coherent or
incoherent way.
 \label{ela}}
\end{figure}

\section{Conclusions}
\label{Conclusions}
The quark-hadron duality in the $\nu$N reactions
has been investigated by comparing the structure functions obtained
from the Rein-Sehgal model and those from the deep inelastic
formalism. The 1-pion functions were used to construct comparable
quantities. The qualitative analysis was based on the plots of the
RS and the DIS structure functions at several values of $Q^2_{RES}$
while the quantitative one was based on the functions
$\mathcal{R}_j$ defined as the ratios of the corresponding
integrals.

We are aware that our model of resonance structure functions is a
subject of several uncertainties. Here is a list of them:

\begin{itemize}

\item[a)] It is possible that the structure functions extracted
from the RS model are underestimated  for $W\geq 1.7$~GeV where
tails of heavier resonances are not included \cite{Bodek-RS}.

\item[b)] In the original RS model the non-resonant background is
treated in not a satisfactory way. In particular the shape of the
non-resonant background does not seem to agree with the precise
electro-production fits presented in Ref. \cite{Backgr}.

\item[c)] Our conclusions depend on the precision with which we
reconstructed 1-pion functions. Therefore we decided also to
investigate the duality with no assumptions about the 1-pion
functions but rather with overall cross sections for the resonance
production (i.e. under the assumption that all the elasticities are
equal 1). We considered two cases: (i) the resonances are added
incoherently and (ii) the interference patterns are the same as in
SPP channels. The typical results are shown in Fig. \ref{ela}:
predictions of all three models turn out to be similar.
\end{itemize}

Our main conclusions can be summarized as follows:
\begin{itemize}

\item[1)] for $Q^2\geq 0.5\,\GeV^2$ the duality is present in the
whole resonance region $W\in (M+m_{\pi}, 2\,\GeV)$ with an accuracy
of $\sim 20\%$ only for CC proton target reaction. We remind that
the way in which the duality is defined carries an uncertainty of
$\sim 5\%$;

\item[2)] from the Figs. \ref{Slizganie_CC} - \ref{slizganie_izo}
it follows that there is also a local duality for isoscalar target
in the case of CC reaction and for all the targets in the case of
NC reactions in a suitable chosen vicinity of the $\Delta$
resonance.

\end{itemize}

The results obtained in this paper can be useful for the
investigation of the question how to modify DIS structure
functions in the low $Q^2$ resonance region so that they provide a
good average description \cite{Reno}. Such modifications should be
confronted with available neutrino scattering data from CHORUS,
NOMAD and NuTeV experiments \cite{Data}.

\section*{Acknowledgements }
\begin{acknowledgements}

The authors (supported by the KBN grant:
105/E-344/SPB/ICARUS/P-03/DZ211/2003-2005) thank Jaros\l aw Nowak
and Olga Lalakulich for helpful conversations.
\end{acknowledgements}


\begin{thebibliography}{99}


\bibitem{BG} E. D. Bloom, F. J. Gilman, Phys. Rev. Lett.
\textbf{25}, 1140 (1970);  Phys. Rev. D \textbf{4}, 2901  (1971).

\bibitem{Nic} I. Niculescu et al., Phys. Rev. Lett. \textbf{85},
1186 (2000).

\bibitem{MEK} For a comprehensive review see: W. Melnitchouk, R. Ent, and C. E. Keppel, Phys.
Rep. \textbf{406}, 127  (2005).

\bibitem{Ru} A. De R\'ujula, H. Georgi, and H. D. Politzer,
Ann. Phys. (N.Y.) {\bf 103}, 315 (1997).

\bibitem{CI} F. E. Close, N. Isgur, Phys. Lett. B {\bf 509},
81 (2001).

\bibitem{CM} F. E. Close, W. Melnitchouk, Phys. Rev. C {\bf 68}, 035210 (2003).

\bibitem{NuInt} Proceedings of the series of NuInt workshops:
Nucl. Phys. B (Proc. Suppl.) {\bf 112} (2002); {\it ibid.} {\bf
139} (2005).

\bibitem{Zel} S. Zeller, {\it Comparisons of available Monte Carlos
with a data}, a talk at {\sl Second International Workshop on
Neutrino-Nucleus Interactions in the Few-GeV Region}, Irvine, Dec
12-15, 2002.

\bibitem{MIN} K. McFarland {\it MINERvA}, a talk at {\sl Fourth International Workshop on
Neutrino-Nucleus Interactions in the Few-GeV Region}, Okayama,
Sept. 26-29, 2005.

\bibitem{RS} D. Rein, L.M. Sehgal, Ann. Phys. \textbf{133}, 79 (1981).

\bibitem{FKR} R.P. Feynman, M. Kislinger, and F. Ravandal, Phys. Rev.
D \textbf{3}, 2706 (1971).

\bibitem{Nau} K.S. Kuzmin, V.V. Lyubushkin, and V.A. Naumov, Mod.
Phys. Lett. A {\bf 19} (2004) 2815.

\bibitem{1PF} C. Juszczak,  J.A. Nowak,  and J.T. Sobczyk, \textit{Simulations
from a new neutrino event generator}, arXiv: hep-ph/0512365.

\bibitem{Fun} S. Mohanty, {\it Treatment of Small Mass Hadronic
Systems in Lund}, a talk at {\sl Second International Workshop on
Neutrino-Nucleus Interactions in the Few-GeV Region}, Irvine, Dec
12-15, 2002.

\bibitem{Sar} F. Sartogo, {\it Interazioni di neutrini atmosferici:
un modelo fenomenologico e la sua applicazione per
l'interpretazione dei dati otenibili in rivelatori soteranei}, PhD
Thesis (in italian), advisor P. Lipari, Rome 1994/95.

\bibitem{Mul} D. Zieminska et al. Phys. Rev. D {\bf 27} (1983) 47;
H. Gr\"assler et al. Nucl. Phys. B {\bf 223} (1983) 269; M.
Arneodo et al. Nucl. Phys. B {\bf 258} (1985) 249; Z. Phys. C {\bf
31} (1986) 1;

\bibitem{Photo} U. Thoma, Int. J. Mod. Phys. A {20} (2005) 1568;
O. Bartholomy et al. Phys. Rev. Lett. {\bf 94} (2005) 012003; S.D.
Ecklund, R.L. Walker, Phys. Rev {\bf 159} (1967) 1195;
T.~A.~Armstrong et al., Phys.\ Rev.\ D {\bf 5} (1972) 1640;
  T.~Fujii et al.,   Nucl.\ Phys.\ B {\bf 120} (1977) 395.




\bibitem{HF} H. Harari, Phys. Rev. Lett. \textbf{20} (1969)
1395; \textit{ibid.} \textbf{22}, 562 (1969); \textit{ibid.}
\textbf{24}, 286 (1970); Ann. Phys. \textbf{63}, 432 (1971); P. G.
O. Freund, Phys. Rev. Lett. \textbf{20}, 235 (1968); P.G.O. Freund
and R.J. Rivers, Phys. Lett. B \textbf{29}, 510 (1969).


\bibitem{Nic2} I. Niculescu et al., Phys. Rev. Lett. \textbf{85}, 1182 (2000).


\bibitem{LS} K. Matsui, T. Sato and T.-S. H. Lee,
Phys. Rev. C \textbf{72}, 25204 (2005).

\bibitem{LP} O. Lalakulich, E. A. Paschos and G. Piranishvili,
{\it Resonance production}, a talk given by O. Lalakulich at {\sl
Fourth International Workshop on Neutrino-Nucleus Interactions in
the Few-GeV Region}, Okayama, Sept. 26-29, 2005.

\bibitem{LS2} T. Sato, D. Uno and T.-S. H. Lee,
Phys. Rev. C \textbf{67}, 065201 (2003).

\bibitem{Lip} P. Lipari, \textit{Calculation of neutrino cross
sections. Open problems. Lines of research}, a talk at {\sl Third
International Workshop on Neutrino-Nucleus Interactions in the
Few-GeV Region}, Gran Sasso, March 17-21, 2004.

\bibitem{Bodek-RS} A. Bodek, \textit{NuInt02 Conference Summary:
Modelling Quasi-elastic, Resonance and Inelastic Neutrino and
Electron Scattering on Nucleons and Nuclei}, a talk at {\sl Second
International Workshop on Neutrino-Nucleus Interactions in the
Few-GeV Region}, Irvine, Dec 12-15, 2002.

\bibitem{Adler} S. Adler, Phys. Rev. \textbf{143} (1966) 1144;
F. Gilman, Phys. Rev. \textbf{167} (1968) 1365.

\bibitem{PDG} The Review of Particle Physics, http://pdg.lbl.gov/ .



\bibitem{Leader} E. Leader, E. Predazzi, \textit{An introduction to gauge
theories and modern particle physics}, vol. I, Cambridge
University Press 1996.


\bibitem{GRV94} M. Gl\"{u}ck, E. Reya, and A. Vogt, Z. Phys. C \textbf{67},
433 (1995).

\bibitem{GJS} K.M. Graczyk, C. Juszczak, and J.T. Sobczyk, \textit{
Appearance of quark-hadron duality in the Rein-Sehgal model},
arXiv: hep-ph/0601077.

\bibitem{BY} A. Bodek, U.K. Yang,
Nucl. Phys. B Proc.(Suppl) {\bf 112}, 70 (2002).

\bibitem{Backgr} S. Galster et al. Phys. Rev. D {\bf 5} (1972)
519.

\bibitem{Reno} M.H. Reno, {\it Electromagnetic structure functions
and neutrino nucleon scattering}, arXiv: hep-ph/0605295.

\bibitem{Data}  G. Onengut et al. Phys.Lett. B. {\bf 632} (2006)
65; R. Petti, {\it Cross-section measurements in the NOMAD
experiment}, arXiv: hep-ex/0602022; NuTeV web page:
http://www-e815.fnal.gov/


\end{thebibliography}
\end{document}